\documentclass[aps,letterpaper,10pt,prb,reprint,nolongbibliography]{revtex4-2}

\usepackage{revquantum}
\usepackage{amsmath}
\usepackage{multirow}
\usepackage{graphics}
\usepackage{graphicx}
\usepackage{times}
\usepackage[table]{xcolor}

\begin{document}

\title{Downfolded Configuration Interaction for Chemically Accurate Electron Correlation}

\author{Jiasheng Li}
\affiliation{Department of Chemistry, The University of Hong Kong,  Hong Kong, China}

\author{Jun Yang}
\email{juny@hku.hk}

\affiliation{Department of Chemistry, The University of Hong Kong,  Hong Kong, China}

\begin{abstract}
A model subspace configuration interaction method is developed to obtain chemically accurate electron correlations by diagonalizing a very compact effective Hamiltonian of realistic molecule. The construction of the effective Hamiltonian is deterministic and implemented by iteratively building a sufficiently small model subspace comprising local clusters of a small number of Slater determinants. Through the low-rank reciprocal of  interaction Hamiltonian, important determinants can be incrementally identified to couple with selected local pairwise clusters and then downfolded into the model subspace. This method avoids direct ordering and selection of the configurations in the entire space. We demonstrate the efficiency and accuracy of this theory for obtaining the near-FCI ground and excited state potential energies by benchmarking C$_2$ molecule, as well as illustrate its application potential in computing accurate excitation energies of organometallic [Cu(NHC)$_2$(pyridine)$_2$]$^{x+}$ complexes and other organic molecules of various excitation character.
\end{abstract}    
    
\maketitle                                        

                                                         
An accurate and efficient determination of correlated electronic structures remains the most important challenge 
in ab-initio computation of atoms and molecules.
The weakly correlated electrons which normally interact within spatially close molecular orbitals can be now accurately 
characterized with high-level methods designed for large molecules,
notably by perturbation or coupled cluster theory and their local correlation variants constructed on a single Hartree-Fock reference wave function. 
However,  stronger correlations in molecule deviating from the equilibrium geometry (bond dissociation, degenerate states, etc.)
tend to spread electrons over a rather large number of orbitals, which continues to prevent practical computations 
from obtaining correct electronic structure using reasonable computing time. 

Strongly correlated electrons have been long attempted by configuration interaction (CI) 
or multireference perturbative computation 
which are conventionally limited to only a small portion of electrons.
Ab-initio density matrix renormalization group (DMRG),
if viewed as a multi-configurational CI technique formulated in the truncated renormalized basis,
has significantly extended the applicability of these approaches for  long molecules with a large number of strongly correlated electrons 
that are effectively one-dimensional \cite{white1992density, white1999ab, chan2002highly, chan2011density}. 
Alternatively, if only a small configurational subspace is important, 
a variety of ideas has been catalyzed to identify important configurations in the full CI (FCI) space, featuring different details and efficiencies.
Important configurations can be stochastically sampled by Quantum Monte Carlo 
\cite{silvestrelli1993auxiliary,zhang2003quantum,thom2010stochastic,booth2009fermion,petruzielo2012semistochastic}, 
or truncated by various selected or adaptive CI methods
 \cite{huron1973iterative, evangelista2014adaptive, holmes2016efficient,
holmes2016heat, tubman2016deterministic,eriksen2017virtual, eriksen2019many, eriksen2020ground, zhang2020iterative, zhang2021further,tubman2020modern, garniron2018selected, loos2020performance, damour2021accurate} 
as well as coupled cluster reduction approach \cite{xu2018full,xu2020towards}.
However, the nature of the exponential growth of the FCI space through which important configurations are examined or sampled is not eliminated,
and the selected subspace can become soon too large to be tamed as the basis sets and the number of correlated electrons are increased.
This leaves a major obstacle for pushing these methods to obtain approximate solution of larger systems. 

In this work, a novel numerical methodology is reported to compute the chemically accurate correlation of many electrons in many orbitals by performing the
direct exact diagonalization in a sufficiently compact model subspace specific to an electronic state, based on Hamiltonian partitioning. 
Using the clusters of determinants exploiting the locality of cluster-specific interactions, a deterministic approach 
is developed to progressively decouple the full Hamiltonian by contracting the many-body effect from the outer subspace onto each cluster of determinants 
of the model subspace,  self-consistently with target energy until convergence.
The iteratively constructed model subspace is  made substantially smaller than a  zero-order space that must be needed in typical quasi-degenerate perturbation 
calculations for chemical accuracy, and suffices a one-to-one mapping between the eigenstates of an intractable full Hamiltonian and the reduced Hamiltonian. 
The method we shall discuss is termed downfolded CI (dCI).

We start from the molecular Hamiltonian $H= \sum_{r,s}^M H_{rs}\ket{\Phi_r}\bra{\Phi_s}$ in the many-electron basis containing  $M$ complete orthogonal determinant functions $\Phi_r$ 
defined in $N$ one-electron molecular orbitals.
Each determinant function $\Phi_r$ is related to a single Slater determinant 
by an excitation manifold and $H_{rs}$ the Hamiltonian element 
with respect to $\Phi_r$ and $\Phi_s$.
Equivalently, the exact energy $E$ of any eigenstate $\Psi$ of $H$ can be mapped to a subspace problem,
\begin{gather}
\overline{H}^P\ket{\Psi_P} = E \ket{\Psi_P},\label{eq:lp1} \\
\Psi = \Psi_P + \frac{\textbf{I}}{E\textbf{I}-QHQ}QHP\Psi_P
\end{gather}
for the effective Hamiltonian $\overline{H}^P$,
\begin{equation}
\overline{H}^P = PHP +PHQ\frac{\textbf{I}}{E\textbf{I}-QHQ}QHP.
\label{eq:lp2}
\end{equation}
This follows from, for instance, the L\"owdin partitioning technique \cite{lowdin1951note, lowdin1962studies}, 
by dividing the full space into the model subspace with the partitioning operator 
$P=\sum_{p\in P}\ket{\Phi_p}\bra{\Phi_p}$
and the outer subspace with $Q=\sum_{q\in Q}\ket{\Phi_q}\bra{\Phi_q}$ 
complementing $P$.
The exact $\overline{H}^P$ has been known practically difficult to obtain 
due to the superoperator $(E\textbf{I}-QHQ)^{-1}$ 
that is large in dimension and nonlinear in an unknown target energy $E$ being sought.
If the coupling $PHQ$ is weak, eq \ref{eq:lp2} can be cast into perturbative expansion of the superoperator 
\cite{lowdin1964studies, lindgren1974rayleigh}. 
For strong couplings, a very large model subspace is normally needed to forge a significant overlap with the strongly correlated target state. 
The effective Hamiltonian partition has been employed to dress a generalized active space CI \cite{li2013splitgas}.
Recently, the model subspace quantum Monte Carlo formalism was proposed to stochastically evaluate molecular $\overline{H}^P$
with quasi-degeneracy \cite{ten2013stochastic, ohtsuka2015study}.

In our method, we set out to divide the model subspace into mutually exclusive clusters $P=\sum_{i} P_i$, 
with each local cluster $P_i$ containing only several quasi-degenerate determinants.
A local quotient term $\mathcal{F}_{ij}$ of $\overline{H}^P$ exists for a pair of clusters $P_i$ and $P_j$ that are algebraically connected to 
the united subspace $Q^{(0)}_{ij}=Q^{(0)}_i\cup Q^{(0)}_j$ using nonzero couplings of $P_{i}HQ^{(0)}_{ij}$ and $P_{j}HQ^{(0)}_{ij}$,
with $Q^{(0)}_i$ and $Q^{(0)}_j$ by at most single and double excitations to $P_i$ and $P_j$, respectively. The corresponding disconnected complements are $\overline{Q}^{(0)}_i=Q-Q^{(0)}_i$ and $\overline{Q}^{(0)}_j=Q-Q^{(0)}_j$.
\begin{subequations}
\begin{eqnarray}
\mathcal{F}_{ij}  =&& \!\!\!\!\!\!\!\!P_{i}HQ\frac{\textbf{I}}{E\textbf{I}-QHQ}QHP_{j} \nonumber \\
=&&\!\!\!\!\!\!\!\! P_{i}HQ^{(0)}_{ij}\left[ R^{(0)}_{ij}+R^{(0)}_{ij} \Sigma^{(0)}_{ij}R^{(0)}_{ij}\right]  Q^{(0)}_{ij}HP_{j} ,\nonumber \\  \label{eq:lp3} \\
\Sigma^{(0)}_{ij} =&&\!\!\!\!\!\!\!\! Q^{(0)}_{ij}H\overline{Q}^{(0)}_{ij}
 \frac{\textbf{I}}{E\textbf{I}-\overline{Q}^{(0)}_{ij}H\overline{Q}^{(0)}_{ij}-V^{(1)}_{ij}} \nonumber \\
  &&\!\!\!\!\!\!\!\! \overline{Q}^{(0)}_{ij}HQ^{(0)}_{ij}, \\
R^{(0)}_{ij}=&&\!\!\!\!\!\!\!\!\frac{\textbf{I}}{E\textbf{I}-Q^{(0)}_{ij}HQ^{(0)}_{ij}},  \\
V^{(1)}_{ij}=&&\!\!\!\!\!\!\!\! \overline{Q}^{(0)}_{ij}HQ^{(0)}_{ij}R^{(0)}_{ij}Q^{(0)}_{ij}H\overline{Q}^{(0)}_{ij}. 
 \label{eq:lp4}
\end{eqnarray} 
\end{subequations}
Eqs \ref{eq:lp3} and \ref{eq:lp4} follow from exact Schur complement based 
on a local superoperator $R^{(0)}_{ij}$ that is invertible in $Q^{(0)}_{ij}$  subspace. 
The remaining disconnected contribution from the primary complement 
$\overline{Q}^{(0)}_{ij}=\overline{Q}^{(0)}_{i}\cup \overline{Q}^{(0)}_j$ 
is important and subsumed in the screened coupling operator 
$\Sigma^{(0)}_{ij}$ which encodes the screening potential $V^{(1)}_{ij}$.

The above construction of $\overline{H}^P$ is exact by summing $P_{i}HP_{j}$ and $\mathcal{F}_{ij}$ 
for all clusters. If the number of clusters is small and the screened coupling operator $\Sigma^{(0)}_{ij}$ 
is known, a simple $\overline{H}^P$ can be built and easily diagonalized.
The practical computation of screened coupling $\Sigma^{(0)}_{ij}$ is therefore critical 
to the determination of $\overline{H}^P$ and hampered by the large complement $\overline{Q}^{(0)}_{ij}$. 
Next,  we adopt a strategy to successively split a cluster-specific complement $\overline{Q}^{(k)}_{i}$  
into connected $Q^{(k+1)}_{i}$ and disconnected complement  
$\overline{Q}^{(k+1)}_{i}$ by nonzero ${Q}^{(k)}_{i}HQ^{(k+1)}_{i}$ ($k=0,1,2,\cdots$) relative to $Q^{(k)}_{i}$, 
\begin{equation} \label{eq:q0}
\overline{Q}^{(k)}_{i}=Q^{(k+1)}_{i}+ \overline{Q}^{(k+1)}_{i},
\end{equation}
where the divisions ($Q^{(0)}_{i},Q^{(1)}_{i},\cdots,Q^{(k)}_{i},\cdots$) 
are mutually exclusive and successively connected, 
and can be viewed as subspace increments towards numerical convergence.
By this construction, the screening $V^{(1)}_{ij}$ survives only in union $Q^{(1)}_{ij}=Q^{(1)}_i\cup Q^{(1)}_j$ connecting $Q^{(0)}_{ij}$,
\begin{equation}
V^{(1)}_{ij}= Q^{(1)}_{ij}HQ^{(0)}_{ij}R^{(0)}_{ij}Q^{(0)}_{ij}HQ^{(1)}_{ij}.
\end{equation}

By applying eq \ref{eq:q0} to $\Sigma^{(0)}_{ij}$,
the generalized screened coupling $\Sigma^{(k)}_{ij}$ is recursively obtained to
downfold many-body interactions onto the subspace $Q^{(k)}_{ij}$ 
in successive connection to higher subspaces ($Q_{ij}^{(k+1)}, Q_{ij}^{(k+2)},\cdots$),
\begin{subequations}
\begin{eqnarray}
\Sigma^{(k)}_{ij}\! =&&\!\!\!\!\!\!\!\! \!Q^{(k)}_{ij}HQ^{(k+1)}_{ij}  \!\!
\left[\!R^{(k+1)}_{ij}\!\!+\!\!R^{(k+1)}_{ij}\Sigma^{(k+1)}_{ij}R^{(k+1)}_{ij}\!\right] \nonumber \\
 \!&&\!\!\!\!\!\!\!\!\! Q^{(k+1)}_{ij}HQ^{(k)}_{ij},  \label{eq:lp5} \\
R^{(k+1)}_{ij}\!=&&\!\!\!\!\!\!\!\!\!\frac{\textbf{I}}{E\textbf{I}-Q^{(k+1)}_{ij}HQ^{(k+1)}_{ij}-V^{(k+1)}_{ij}},  \label{eq:lp6}\\
V^{(k+1)}_{ij}\!=&&\!\!\!\!\!\!\!\!\!Q^{(k+1)}_{ij}HQ^{(k)}_{ij}R^{(k)}_{ij}Q^{(k)}_{ij}HQ^{(k+1)}_{ij}. \label{eq:lp7}
\end{eqnarray}
\end{subequations}
The superoperator $R^{(k+1)}_{ij}$, which extends $R^{(0)}_{ij}$, is local to the cluster pair $ij$ and dynamically screened by the generalized potential $V_{ij}^{(k+1)}$.
As this recursion in eq \ref{eq:lp5} suggests, a subspace increment
$Q^{(k+1)}_{ij}$ necessitates the revision of all prior screened coupling operators 
by backpropagating
$\Sigma^{(k)}_{ij}\rightarrow \Sigma^{(k-1)}_{ij}\cdots \rightarrow \Sigma^{(0)}_{ij}$, 
which in turn updates $\mathcal{F}_{ij}$ over all local clusters 
and eventually $\overline{H}^P$. 

\begin{figure}[htbp]
\begin{center}
\includegraphics[width=9cm]{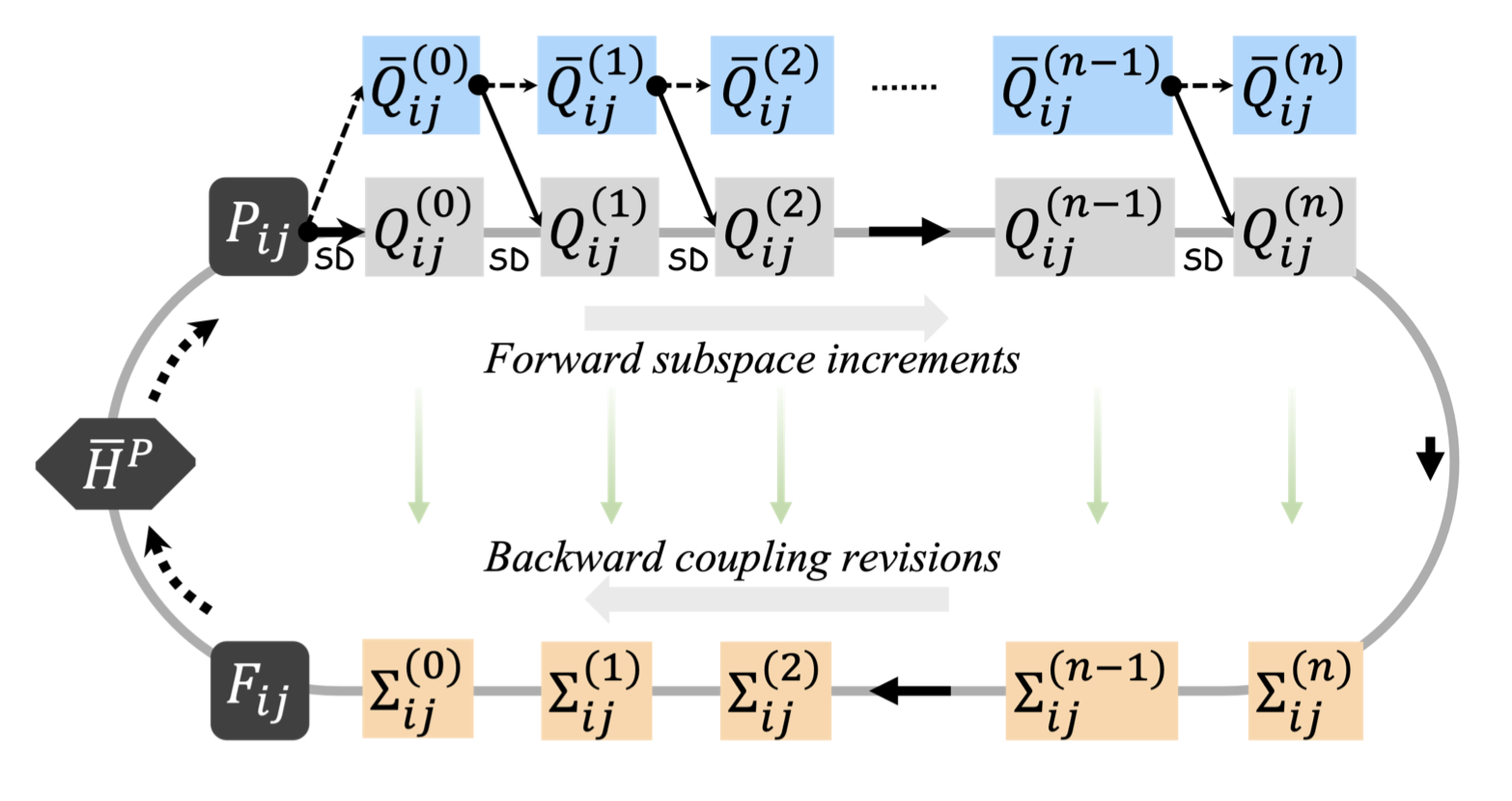}
\caption{The schematic illustration of the iterative dCI method and the construction of the model 
and outer subspaces. Each increment of the outer subspace is connected to prior subspace 
by single and double (SD) excitation manifolds.}
\label{fig:scheme}
\end{center}
\end{figure}

Eqs \ref{eq:lp5} and \ref{eq:lp7} reveal the following compelling criteria for identifying important determinants.
(i) By utilizing the subspace hierarchy, the selection of important determinants belonging to $Q_{ij}^{(k+1)}$ 
can be narrowed to the local subspace connected to prior $Q_{ij}^{(k)}$ only for each $ij$ pair at a time; 
and (ii) by preordering molecular integrals that attenuate the coupling  magnitudes 
${Q^{(k)}_{ij}HQ^{(k+1)}_{ij}}$,  
the recursion can be converged quadratically in the coupling strength as the product
$||{Q^{(k)}_{ij}HQ^{(k+1)}_{ij}}||\cdot||Q^{(k+1)}_{ij}HQ^{(k)}_{ij}||$ arises.  
The combination of (i) and (ii) and the proper implementation of them enable low-rank local superoperators ($R^{(0)}_{ij}, R^{(1)}_{ij}, \cdots$) 
and an efficient backpropagating determination of local screened coupling operators 
($\Sigma^{(0)}_{ij}, \Sigma^{(1)}_{ij},\cdots$) upon the forward subspace increment, 
as illustrated in Figure \ref{fig:scheme}.
Therefore, an enormous exploration of the  FCI basis is unnecessary.

The most salient feature of dCI is a configuration selection by the quadratic outer coupling in $Q_{ij}^{(k)}HQ_{ij}^{(k+1)}$,  as opposed to the linear coupling in other selected CI methods. The dCI search for important determinants is constrained within incremental subspaces successively connected to a small local cluster at a time, and a simple parallelization scheme performing dCI operations over separate local clusters can be envisaged.

For a pair of clusters $P_{ij}$, its fully connected complement $Q^{(0)}_{ij}$  is in practice numerically pruned to reduce the operations described above.
Only the determinants $\Phi^{(0)}_{q}\in Q^{(0)}_{ij}$ that couple significantly to  all cluster determinants $\Phi_p \in P_{ij}$  
are selected by the comparison $\sum_p |\bra{\Phi_{p}}H\ket{\Phi^{(0)}_{q}}|>\theta_{I_1}$  using the first integral cutoff $\theta_{I_1}$. 
The determinants $\Phi^{(k)}_q$ in higher $Q^{(k)}_{ij}$ subspace increments are discriminated based on the local
screening potential $|\bra{\Phi_{q}^{(k)}}V^{(k)}_{ij}\ket{\Phi_{q}^{(k)}}|>\theta_{I_2}$ 
by comparing to the second integral cutoff $\theta_{I_2}$. For all present calculations, we found it sufficient 
to use $\theta_{I_1}=10^{-8}$ and $\theta_{I_2}=10^{-6}$.

\begin{figure}[ht!]
\begin{center}
\includegraphics[trim={1cm 0 0 0},clip,width=8cm]{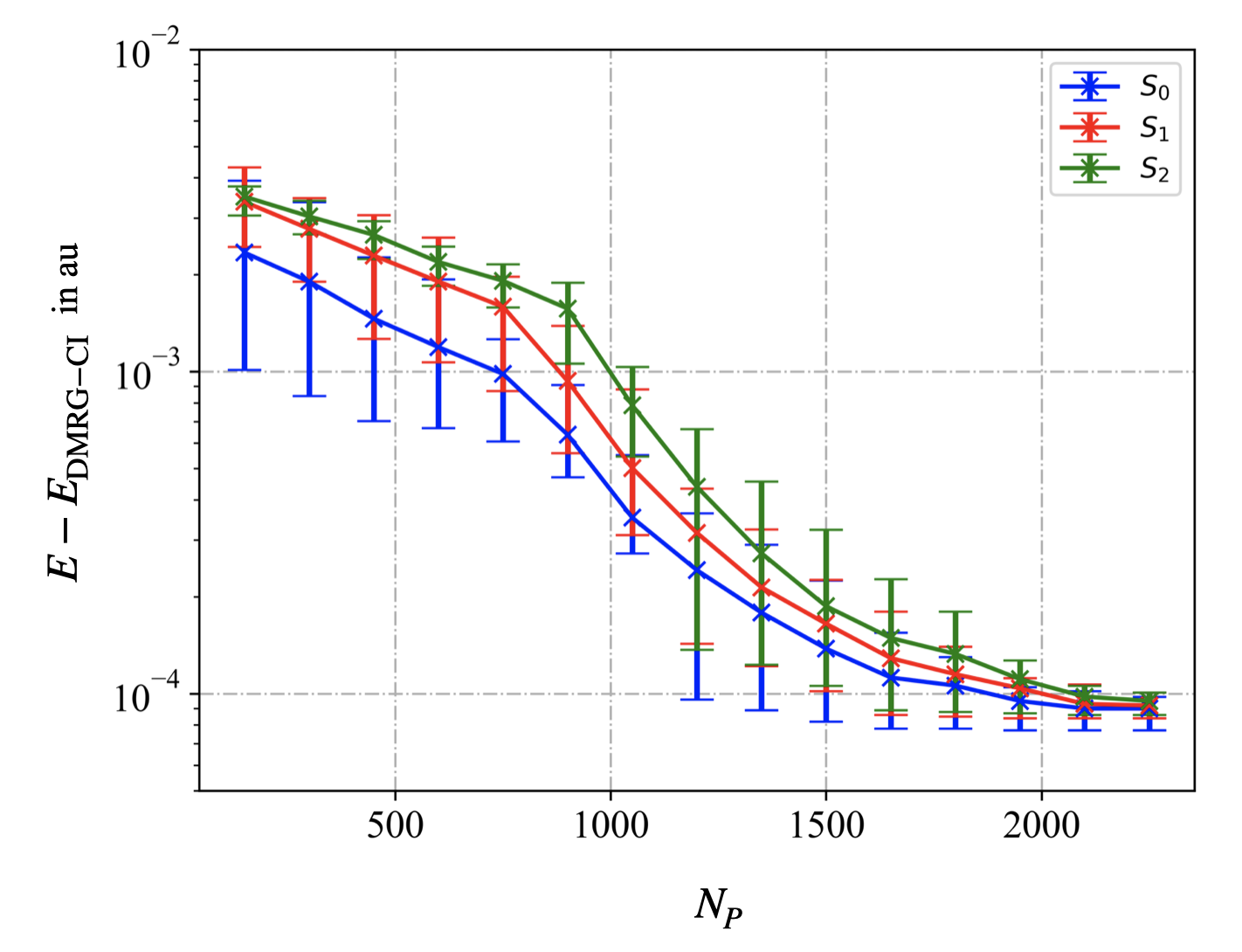}
\caption{Convergence of the dCI energy errors for S$_n$ ($n=0,1,2$) singlet states  of C$_2$ with respect to the model subspace dimension $N_P$. The dCI iteration is converged by $\theta_E=10^{-4}~\text{au}$ in the cc-pVTZ basis. All 12 electrons are correlated in the complete 60 Pipek-Mezey local orbitals. The error bar corresponds to the range by the minimal and maximal absolute energy deviations among all nine C$_2$ structures ($d_\mathrm{C-C}$=1.0, 1.2, $\cdots$, 2.6 \AA).}
\label{fig:fig1}
\end{center}
\end{figure}
\begin{figure}[ht!]
\begin{center}
\includegraphics[trim={1cm 0 0 0},clip,width=8cm]{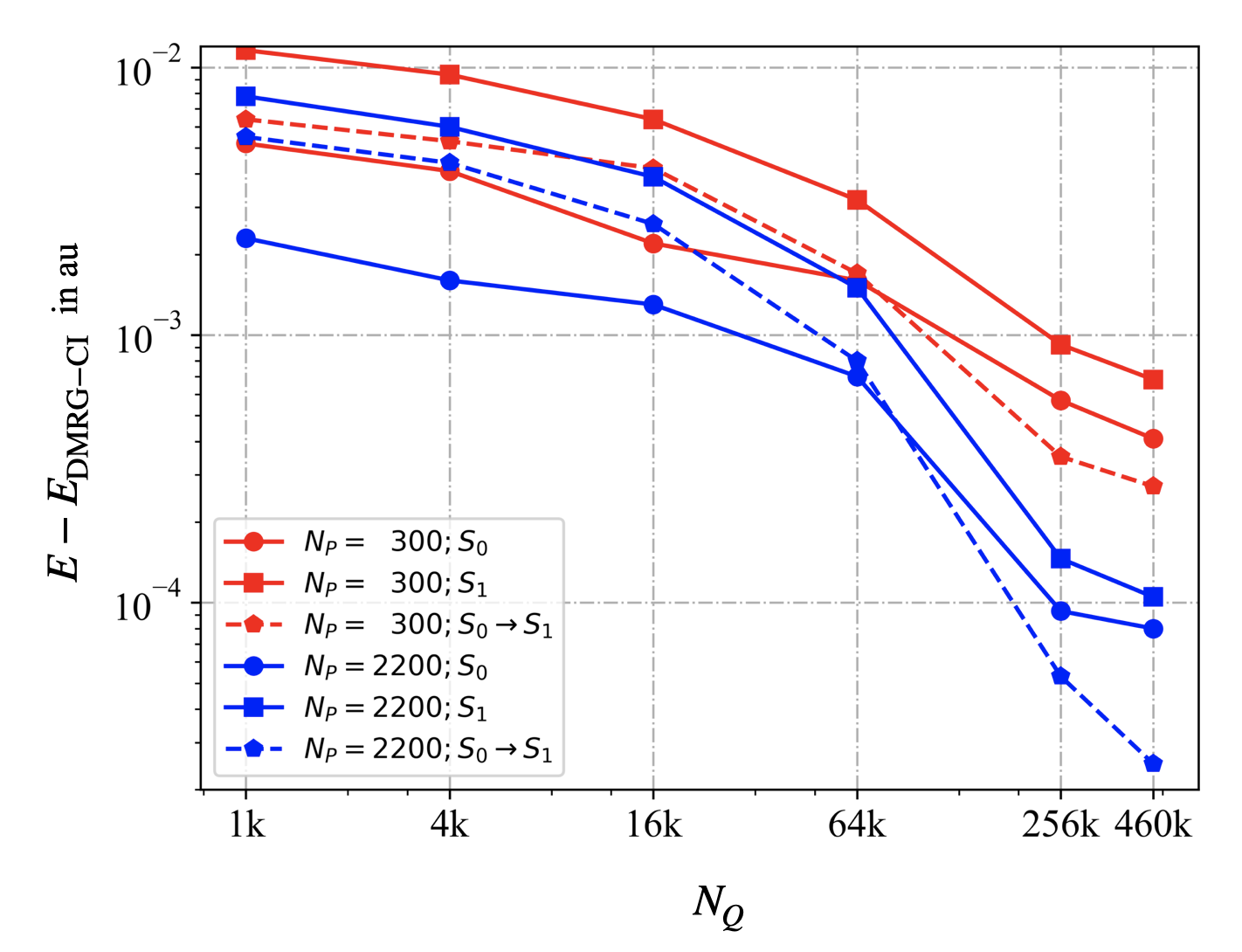}
\caption{Convergence of the dCI energy errors for S$_0$ ($X {\ }^1\Sigma_g^+$) and S$_1$ ($B {\ }^1\Delta_g$) states of C$_2$ ($d_\mathrm{C-C}$=1.24253 \AA) with respect to the outer subspace dimension $N_Q$ in the cc-pVTZ basis.}
\label{fig:fig2}
\end{center}
\end{figure}

\begin{table*}[ht!]
\footnotesize
\caption{The error comparison of singlet excitation energies ($\Delta\omega$) between dCI and DMRG-CI ($M=4000$) results with respect to the dimensions of full ($N_\text{FCI}$), model ($N_P$) and outer ($N_Q$) spaces for organic molecules. All valence electrons are correlated in all molecular orbitals in the cc-pVTZ basis (6-31g for charge transfer excitations of NH$_3\rightarrow$F$_2$ and N$_2\rightarrow$CH$_2$).}
\begin{center}
\begin{tabular}{cccccccccccc}
\hline\hline
\multirow{2}{*}{Molecules} &\multirow{2}{*}{Characters}&  \multirow{2}{*}{Active Space } & \multirow{2}{*}{$N_\text{FCI}$} &  \multicolumn{3}{c}{$N_P(N_Q)$$^a$ ($\times10^3$)} & \multicolumn{3}{c}{$\Delta\omega$$^b$ (eV)} \\ 
\cline{5-7}\cline{8-10} 
&&&&$\theta_\phi^l$ & $\theta_\phi^m$  & $\theta_\phi^t$ & $\theta_\phi^l$ & $\theta_\phi^m$ & $\theta_\phi^t$   \\
\hline
NH$_3\rightarrow$F$_2$                    & charge transfer                                &    (30o, 22e) &$3.0\times 10^{15}$ &    0.31(50) & 0.51 (100) & 0.91 (400) & 0.26 & 0.18 &0.049\\

N$_2\rightarrow$CH$_2$                 & charge transfer                                  &    (28o, 16e) &$9.7\times 10^{12}$ &     0.16(50) & 0.36 (100) & 0.67 (250)& 0.19 & 0.11 &0.038  \\

C$_2$H$_4$                                      & $1^1 A_g \rightarrow 2^1 A_g$       &    (114o, 12e) &$7.1\times 10^{18}$  &  0.38 (100) & 0.59 (200) & 1.22 (550) &  0.27 & 0.19 &0.052\\

C$_3$                               & $1^1 \Sigma_g^+ \rightarrow 1^1 \Delta_g$      &   (87o, 12e) &$2.6\times 10^{17}$ & 0.32 (100) & 0.52 (150) & 0.95 (450) &  0.32 & 0.24 &0.046 \\

HNO                                       &  $1^1 A' \rightarrow 2^1 A'$                          &    (72o, 12e) &$2.4\times 10^{16}$&0.28 (50) & 0.46 (100) & 0.82 (350) &  0.23 & 0.17 &0.042 \\

H$_2$S                         &  $S_0 \rightarrow S_1$                                            &   (61o,  16e)  &$8.7\times 10^{18}$  & 0.42 (100) & 0.65 (200) & 1.14 (500) &  0.28 & 0.23 &0.041\\

HCHO                             &  $S_0 \rightarrow S_1$                                          &     (86o, 12e) & $2.2\times 10^{17}$& 0.35 (100) & 0.55 (150) & 0.98 (450)&  0.33 & 0.20 &0.053 \\
\hline\hline  
\multicolumn{10}{l}{$^a$ Various dCI overlap cutoffs: $\theta_\phi^l=10^{-4}$ (loose),  $\theta_\phi^m=10^{-5}$ (medium) and  $\theta_\phi^t=10^{-6}$ (tight).} \\ 
\multicolumn{10}{l}{$^b$ The dCI energy threshold: $\theta_E=10^{-4}$ au.} 
\end{tabular}\label{tab:tab1}
\end{center}
\end{table*}%

The trial model subspace $P$ can be composed of a few Slater determinants 
according to the magnitudes of their CI coefficients derived from a complete active space CI (CAS-CI) wave function. However,
a projected Hamiltonian $PHP$ using a poor  model subspace $P$ may not have an eigenspectrum significantly overlapping the target state,
which results in slow numerical convergence. This problem is circumvented by
incorporating the target state character in a delicate self-bootstrapping procedure, as detailed below.
A new  cluster $P_{i'}$ is generated by moving out several spin-adapted quasi-degenerate determinants 
$\Phi'^{(0)}_q \in Q_{i}^{(0)}$ which possess evident energy differences from the existing cluster $P_i$
and significant overlaps with the proceeding target root $\ket{\Psi_P}$ (or multiple degenerate roots) 
of $\overline{H}^P$ using the overlap cutoff $\theta_\phi$,
\begin{equation}
\bra{\Phi'^{(0)}_q}\sum_{i,j}[R^{(0)}_{ij}+R^{(0)}_{ij} \Sigma^{(0)}_{ij}R^{(0)}_{ij} ]Q^{(0)}_{ij}HP_{ij}\ket{\Psi_P}> \theta_\phi. \label{eq:thetap}
\end{equation} 
The corresponding complementary outer $Q^{(0)}_{i'}$ is simply formed by sequentially shifting selected determinants along the path 
$Q^{(k+1)}_{i}\rightarrow Q^{(k)}_{i'}$ which automatically follows the connectivities in eq \ref{eq:q0}.
The model subspace update is repeated until self-consistent change of the target state energy $\delta E$ meeting the threshold $|\delta E| < \theta_E$.
\begin{table*}[ht!]
\caption{Comparison of excitation energy ($\omega$) and CI dimension between dCI ($\theta_\phi^t$=$10^{-6}$) and DMRG-CI ($M=4000$) results for [Cu(NHC)$_2$(pyridine)$_2$]$^{x+}$ (x=1, 2, 3) complexes of different Cu oxidation state.}  
\begin{center}
\begin{tabular}{cccccccccc}
\hline\hline
  & \multirow{2}{*}{States} & \multirow{2}{*}{$N_\text{FCI}$ ($\times 10^{16}$)} & \multirow{2}{*}{$N_P$ ($\times 10^3$ )} & \multirow{2}{*}{$N_Q$ ($\times 10^3$)} &  \multicolumn{3}{c}{$\omega$ (eV)}\\ 
\cline{6-8} & & & &  & dCI & &DMRG-CI  && \\
\hline
[CuN$_6$C$_{20}$H$_{18}$]$^+$        &   $\text{S}_1$  & 2.4 & 1.62 & 440 & 6.32 && 6.28  &&\\
                                                                    & $\text{S}_2$    &       & 1.73 & 470 & 6.81 && 6.77 &&\\ [0.8mm]                                                  
[CuN$_7$C$_{22}$H$_{21}$]$^{2+}$   & $\text{D}_1$  &  2.3 & 1.45 & 440 & 5.24 && 5.19 &&\\ 
                                                                    & $\text{D}_2$  &        & 1.75 & 460 & 5.29 && 5.24 &&\\   [0.8mm]                                                      
[CuN$_7$C$_{22}$H$_{21}$]$^{3+}$   & $\text{S}_1$  & 2.4  & 1.64 & 480  & 4.43 &&4.38 && \\
                                                                    & $\text{S}_2$  &         & 1.84 & 470 & 4.35 && 4.31 &&\\                                                         
\hline\hline 
\multicolumn{5}{l}{NHC: N-heterocyclic carbene ligand.}
\end{tabular}\label{tab:tab2}
\end{center}
\end{table*}%

We assess dCI performance concerning the convergence routes of Figure \ref{fig:scheme} with respect to the dimension of both model and outer subspaces, and the results are shown in Figures \ref{fig:fig1} and \ref{fig:fig2} for C$_2$ molecule using cc-pVTZ basis. The dCI energy error relative to the DMRG-CI benchmark is reduced by increasing the number of determinants ($N_P$) in the model subspace, and falls within $10^{-3}$ and $10^{-4}$ au using $N_P$ = $1000$-$1200$ and $N_P$ = $1800$-$2300$, respectively, as shown in Figure \ref{fig:fig1}. 
Although the original FCI space of $C_2$ molecules requires about $\sim10^{15}$ determinants by varying C-C bond length, Figure \ref{fig:fig1} illustrates that it is possible to obtain chemically accurate (and near-FCI) correlation energies in a rather compressed model subspace in which an effective Hamiltonian on can be found and simply diagonalized. 

We further demonstrate the convergence of the state and excitation energies with respect to the dimension of the outer subspace by fixing $N_P$ for C$_2$/cc-pVTZ in Figure \ref{fig:fig2}. The dCI correlation energies are continuously improved by incorporating the essential determinants in the complementary subspace and approach chemical (~$\sim1$ kcal/mol) and near-FCI ($\sim0.05$ kcal/mol) accuracy in a rather compact $N_Q$=64-100 k and $N_Q$=460 k, respectively, which is downfolded into the effective Hamiltonian of the dimension by $1200\times1200$ that is much smaller than in the original FCI dimension for direct diagonalization. Moreover, the vertical $\mathrm{S}_0\rightarrow\mathrm{S}_1$ and $\mathrm{S}_0\rightarrow\mathrm{S}_2$  excitation energies appear converged within 0.01 eV, requiring only $N_Q$=100 k determinants for $N_P=2200$ and greater $N_Q$=250 k for smaller $N_P$=300, both of which are even much milder than those for total energies.  These numerical results demonstrate that our dCI algorithm leads to an automatic tradeoff between the construction of the model and outer subspaces.

More examples of molecular excitation energies for a few low-lying singlets of different characters in organic molecules are compared to the reference DMRG-CI benchmark. These molecules yield a prohibitive FCI expansion for the practical cc-pVTZ basis set. Table~\ref{tab:tab1} demonstrates the dCI convergence of excitation energies with respect to various overlap cutoffs that determine the complementary subspace according to eq \ref{eq:thetap}. For small errors of $\Delta \omega$=0.04-0.05 eV, the tight cutoff  is necessary and produces only $N_P$=650-1200  and $N_Q$=200-550~k determinants. These results clearly reveal the advantage of dCI method: despite of the diverse multireference character of these molecules, as reflected by the broad range of FCI determinants from  nearly $10^{13}$ to $10^{19}$, the resulting sizes of dCI model and outer subspaces differ only by less than two folds. 

Finally, we show an illustrative application of the dCI method to excitation energies between a few singlet and doublet states for organometallic Cu(I)/Cu(II)/Cu(III) complexes coordinated to double NHC and pyridine ligands which are of more chemical interests. Using moderately large def2-TZVP basis set, occupied and virtual molecular orbitals were localized separately by Pipek-Mezey method. The active space contains 30 molecular orbitals for 30 or 29 valence electrons for valency $x=1, 3$ and $x=2$, respectively, with predominant atomic orbitals of Cu/3d, C/2p and N/2p. The comparison of the subspace sizes and energies is shown in Table~ \ref{tab:tab2}, and indicates that the original CI problem of the prohibitive dimension by $10^{16}$ can be downfolded into a highly compressed dCI Hamiltonian.

To summarize, we introduced the dCI algorithm for ab-initio electronic structure calculation by efficient many-body partitioning. The cluster localization and the deterministic-based truncation of the model subspace break the exponential scaling as encountered in the conventional configuration interaction scheme. We showed the dCI capability in obtaining chemically accurate  energy properties of various molecular systems, highlighting a robust and systematic numerical approach to find a sufficiently compact Hamiltonian partitioning. In particular, the original problem in the FCI expansion at orders of nearly $10^{13}$-$10^{19}$ basis can be successfully tackled by contracting only most essential determinants ($N_Q\sim~10^{5}$) into a small subspace ($N_P$=$500$-$2000$), probably due to more sparsity of the screened coupling operator than the FCI Hamiltonian. Therefore, dCI appears to be a promising scheme for treating correlated molecules with balanced model and outer subspaces. 
Extensions of the dCI approach can be explored in different ways. For instances, dCI provides a way to examine the possibility of constructing numerically exact many-body quantum embedding basis for molecules; the pairwise nature of dCI algorithm advises to implement efficient parallelism for larger molecules; orbital relaxations and perturbation correction may further trim the Hamiltonian size; the multistate dCI method is also underway.

\section*{Acknowledgements}
Financial supports from Hong Kong Research Grant Council (GRF17309020 and
GRF17310922) are acknowledged. We are grateful to Hong Kong Quantum AI Lab Ltd
and the Computational Initiative Program by the Faculty of Science at The
University of Hong Kong.

\section*{Supporting Information}

The Supporting information contains further details, results and molecular geometries
for the computations and is available free of charge.

\section*{Code Availability}

The sample dCI program is available at {https://github.com/QCLabHKU/dci-demo}. 


\bibliography{manuscript}



\end{document}